\documentclass[preprint2]{aastex}
\usepackage{url}

\slugcomment{}

\shorttitle{A Young Massive Stellar Population Around ESO 43-49 HLX-1}
\shortauthors{Farrell et al.}

\begin{document}


\title{A Young Massive Stellar Population Around the Intermediate Mass Black Hole ESO 243-49 HLX-1}

\author{S. A. Farrell\altaffilmark{1,2}, M. Servillat\altaffilmark{3}, J. Pforr\altaffilmark{4,5}, T. J. Maccarone\altaffilmark{6}, C. Knigge\altaffilmark{6}, O. Godet\altaffilmark{7,8}, C. Maraston\altaffilmark{4}, N. A. Webb\altaffilmark{7,8}, D. Barret\altaffilmark{7,8}, A. J. Gosling\altaffilmark{9}, R. Belmont\altaffilmark{7,8}, and K. Wiersema\altaffilmark{2}}
\email{sean.farrell@sydney.edu.au}

\altaffiltext{1}{Sydney Institute for Astronomy (SIfA), School of Physics, The University of Sydney, NSW 2006, Australia}

\altaffiltext{2}{Department of Physics and Astronomy, University of Leicester, University Road, LE1 7RH, Leicester, UK}

\altaffiltext{3}{Harvard-Smithsonian Center for Astrophysics, 60 Garden Street, MS-67, Cambridge, MA 02138, USA}
\altaffiltext{4}{Institute of Cosmology and Gravitation, University of Portsmouth, Dennis Sciama Building, Burnaby Road, Portsmouth, PO1 3FX, UK}

\altaffiltext{5}{Present address: National Optical Astronomy Observatory (NOAO), 950 North Cherry Avenue, Tuscon, AZ 85719, USA}

\altaffiltext{6}{School of Physics and Astronomy, University of Southampton, Hampshire, SO17 1BJ, UK}

\altaffiltext{7}{Universit\'{e} de Toulouse, UPS-OMP, Institut de Recherche en Astrophysique et Plan\'{e}tologie (IRAP), Toulouse, France}

\altaffiltext{8}{CNRS, IRAP, 9 Avenue du Colonel Roche, BP 44346, F-31028 Toulouse Cedex 4, France}

\altaffiltext{9}{University of Oxford, Department of Physics, Keble Road, Oxford, OX1 3RH, UK}



\begin{abstract}

We present \emph{Hubble Space Telescope} and simultaneous \emph{Swift} X-ray telescope observations of the strongest candidate intermediate mass black hole (IMBH) ESO 243-49 HLX-1. Fitting the spectral energy distribution from X-ray to near-infrared wavelengths showed that the broadband spectrum is not consistent with simple and irradiated disc models, but is well described by a model comprised of an irradiated accretion disc plus a $\sim$10$^6$ M$_\odot$ stellar population. The age of the population cannot be uniquely constrained, with both young and old stellar populations allowed. However, the old solution requires excessive disc reprocessing and an extremely small disc, so we favour the young solution ($\sim$13 Myr). In addition, the presence of dust lanes and the lack of any nuclear activity from X-ray observations of the host galaxy suggest that a gas-rich minor merger may have taken place less than $\sim$200 Myr ago. Such a merger event would explain the presence of the IMBH and the young stellar population.

\end{abstract}





\section{Introduction}

The formation of stellar-mass black holes (BHs) \citep[$\sim$3 -- 80 M$_\odot$;][]{bel10} through the collapse of massive stars is well accepted, but it is not yet completely clear how the supermassive BHs ($\sim$10$^6$ -- 10$^9$ M$_\odot$) are formed. They may form through the merger of $\sim$10$^2$ -- 10$^5$ M$_\odot$ IMBHs  \citep{ebi01}. IMBHs are thus a crucial missing link between stellar-mass and supermassive BHs, with likely environments for their formation including globular clusters \citep{mil02} and the nuclei of dwarf galaxies \citep{ric04}. The existence of IMBHs also has implications for other areas of astrophysics including the search for dark matter annihilation signals \citep{for08}, the epoch of reionization of the Universe \citep{ric04}, and the detection of gravitational wave radiation \citep{abb09}. The study of IMBHs and the environments in which they are found thus has important connotations for a wide range of important questions in modern astrophysics.

The brightest ultra-luminous X-ray source ESO 243-49 HLX-1 currently provides the strongest evidence for the existence of IMBHs \citep{far09}. HLX-1 is located in the halo of the edge-on S0a galaxy ESO 243-49, $\sim$0.8 kpc out of the plane and $\sim$3.3 kpc away from the nucleus. At the redshift of ESO 243-49 (z = 0.0223) the maximum 0.2 -- 10 keV X-ray luminosity of HLX-1 is $\sim$1.3 $\times$ 10$^{42}$ erg s$^{-1}$ \citep{god11a}, a factor of $\sim$400 above the theoretical Eddington limit for a 20 M$_\odot$ BH. Luminosities up to $\sim$10$^{41}$ erg s$^{-1}$ can be explained by stellar-mass BHs undergoing super-Eddington accretion \citep{beg02} and/or experiencing significant beaming, which makes them appear to exceed the Eddington limit for isotropic radiation \citep{kin08,fre06}. However, luminosities above $\sim$10$^{41}$ erg s$^{-1}$ are difficult to explain without a more massive BH. Following the discovery of an optical counterpart by \citet{sor10}, the distance to HLX-1 was confirmed through the detection of the H$\alpha$ emission line at a redshift consistent with the host galaxy \citep{wie10}, confirming the extreme luminosity. Modelling the accretion disc emission using relativistic \citep{dav11} and slim disc \citep{god11b} models implies a mass between $\sim$3,000 -- 100,000 M$_\odot$, consistent with values obtained by scaling from stellar-mass BH binaries \citep{sev11}. 


Long-term monitoring with the \emph{Swift} observatory has shown that HLX-1 varies in X-ray luminosity by a factor of $\sim$50 \citep{god09}, with correlated spectral variability reminiscent of that seen in Galactic stellar-mass BHs \citep{sev11}. Since the \emph{Swift} monitoring began in 2009, HLX-1 has been observed to undergo three outbursts, each spaced approximately one year apart  \citep{god11b}. The characteristic timescales of the outbursts are inconsistent with the  thermal-viscous instability model, and the outburst mechanism could instead be tidal stripping of a companion star in an eccentric orbit \citep{las11}.

Excess UV emission was detected consistent with HLX-1 with the \emph{GALEX} and \emph{Swift} observatories, although this emission could not be resolved from the nucleus of ESO 243-49 \citep{web10}. At least some of the UV excess is likely to be associated with a z $\sim$ 0.03 background galaxy \citep{wie10,far11}. Using preliminary UV fluxes obtained from the \emph{Hubble Space Telescope} (\emph{HST}) data presented in this paper, \citet{las11} estimated the age of the stellar environment around HLX-1 to be $\sim$0.3 -- 0.6 Gyr  through stellar population synthesis modelling and the assumption of a 5 $\times$ 10$^6$ M$_\odot$ cluster mass and Solar metallicities. They therefore concluded that the most likely progenitor of the companion star would be an AGB star with a mass $\sim$2.7 -- 3.5 M$_\odot$. However, the contribution from accretion disc emission (in particular possible irradiation in the outer disc) were not taken into account, and so a more rigorous approach employing broadband modelling is required.

In this paper we report the results of detailed broadband spectral modelling of HLX-1 using UV, optical and near-IR observations performed with the \emph{HST} in conjunction with simultaneous observations performed in X-ray wavelengths with the \emph{Swift} X-ray Telescope (XRT). The paper is organised as follows: \S 2 lists the data reduction steps, while \S 3 describes the analysis methods employed and the results obtained. \S 4 discusses the implication of these results and the conclusions that we have drawn. 

\section{Data Reduction}



Following the peak of the second outburst we obtained three orbits of observations with the \emph{HST} on  2010 September 13 and 23 under program \#12256 in order to constrain the nature of the environment around HLX-1. During these observations HLX-1 was in the thermally dominated high/soft spectral state. Observations were performed in the far-UV (FUV) band using the Advanced Camera for Surveys (ACS) Solar Blind Camera (SBC), and in the near-UV (NUV), Washington C, V, I and H bands with the Wide Field Camera 3 (WFC3) UVIS and IR cameras. Table \ref{astro}  presents the log of the \emph{HST} observations. We analyzed the final \emph{HST} images generated by the pipeline (drz files) using the latest calibration data (CALWF3=2.1 as of 2010 May 15, and CALACS=5.1.1 as of 2010 April 27). As part of the pipeline these images were flat fielded, combined (with cosmic ray rejection) and geometrically corrected using  {\tt PyDrizzle} v6.3.5 (2010 May 19). 

We corrected the astrometry of each \emph{HST} image following a two-step procedure with an intermediate image and using the US Naval Observatory CCD Astrograph Catalog \citep[UCAC3;][]{zac10} as an astrometric reference (typical rms error of 0.1\arcsec). As the \emph{HST} WFC3 and ACS fields of view are relatively small and the target is located in a particularly empty field that only contains 1 or 2 UCAC3 sources (including the extended source ESO 243-49 itself), we used an intermediate image obtained with the Infrared Side Port Imager (ISPI) at the Cerro Tololo inter-American observatory \citep[CTIO;][]{van04} on 2009 August 2. This image covers a 10\arcmin $\times$10\arcmin\ region around ESO 243-49 in the J-band. We aligned the ISPI image on the UCAC3 catalogue using the  {\tt Starlink}/ {\tt GAIA} software and obtained a precision of 0.13\arcsec\ (rms) using 9 reference stars. We then used the ISPI image as a reference to align the WFC3 images. The number of stars used in the process and the resulting absolute position error are given in Table \ref{astro} for each band. For the ACS-SBC image (F140LP filter), this process could not be applied because of the lack of detection of ISPI stars in this band. We thus used the WFC3-UVIS near-UV image as a reference, finding 4 common objects. The precision is also reported in Table \ref{astro}. The HLX-1 \emph{Chandra} X-ray position is 0.3\arcsec\ at 95$\%$ confidence \citep{web10}. The \emph{HST} errors include the UCAC3 absolute error of 0.05\arcsec\ and the ISPI relative error of 0.13\arcsec\ (1$\sigma$). The final position error, adding in quadrature all errors, is thus 0.5\arcsec\ at 95\% confidence level for each \emph{HST} image. A single counterpart is clearly detected in each band within this error circle (Figure \ref{altimg}), although in the F160W H-band filter image it is difficult by eye to see the counterpart against the strong diffuse emission from the galaxy. 

\begin{table*}
\begin{center}
\caption{Log of the \emph{HST} observations and details of the astrometric corrections applied to the images.\label{astro}}
\begin{tabular}{cccccccc}
\tableline\tableline
Instrument & Filter & Date & t$_{exp}$ & N$_{ISPI}$\tablenotemark{a} & Fit rms & \multicolumn{2}{c}{Astrometric Error\tablenotemark{b}} \\ 
 & & &(s) &&& \emph{HST} & Total \\
\tableline
ACS-SBC & F140LP & 2010-09-13 & 2480 & 4\tablenotemark{c} & 0.01\arcsec & 0.35\arcsec & 0.47\arcsec \\
WFC3-UVIS & F300X & 2010-09-23 & 1710 & 6 & 0.05\arcsec & 0.35\arcsec & 0.47\arcsec \\
WFC3-UVIS & F390W & 2010-09-23 & 712 & 8 & 0.05\arcsec & 0.35\arcsec & 0.47\arcsec \\
WFC3-UVIS &  F555W & 2010-09-23 & 742 &18 & 0.09\arcsec & 0.40\arcsec & 0.50\arcsec \\
WFC3-UVIS & F775W & 2010-09-23 & 740 & 25 & 0.09\arcsec & 0.40\arcsec & 0.50\arcsec \\
WFC3-IR & F160W & 2010-09-23 & 806 & 22 & 0.10\arcsec & 0.41\arcsec & 0.50\arcsec \\
\tableline
\end{tabular}
\tablenotetext{a}{Number of ISPI sources used in the correction.}
\tablenotetext{b}{All errors quoted are at the 95\%\ significance level.}
\tablenotetext{c}{The F140LP image was aligned on the F300X image, and not on ISPI directly.}

\end{center}
\end{table*}

The field of HLX-1 was also observed with the \emph{Swift} XRT (obsid's: 00031287055, 00031287056, 00031287057, 00031287058, and 00031287059) on 2010 September 13, 14, and 23 for a total of 17.5 ks. The data were reduced, and spectra were extracted in the same manner as described in \citet{sev11}. The spectra were binned to a minimum of 20 counts per bin in order to use $\chi^2$ statistics for the fitting.

\section{Data Analysis \& Results}

\begin{figure*}
\epsscale{2}
\plotone{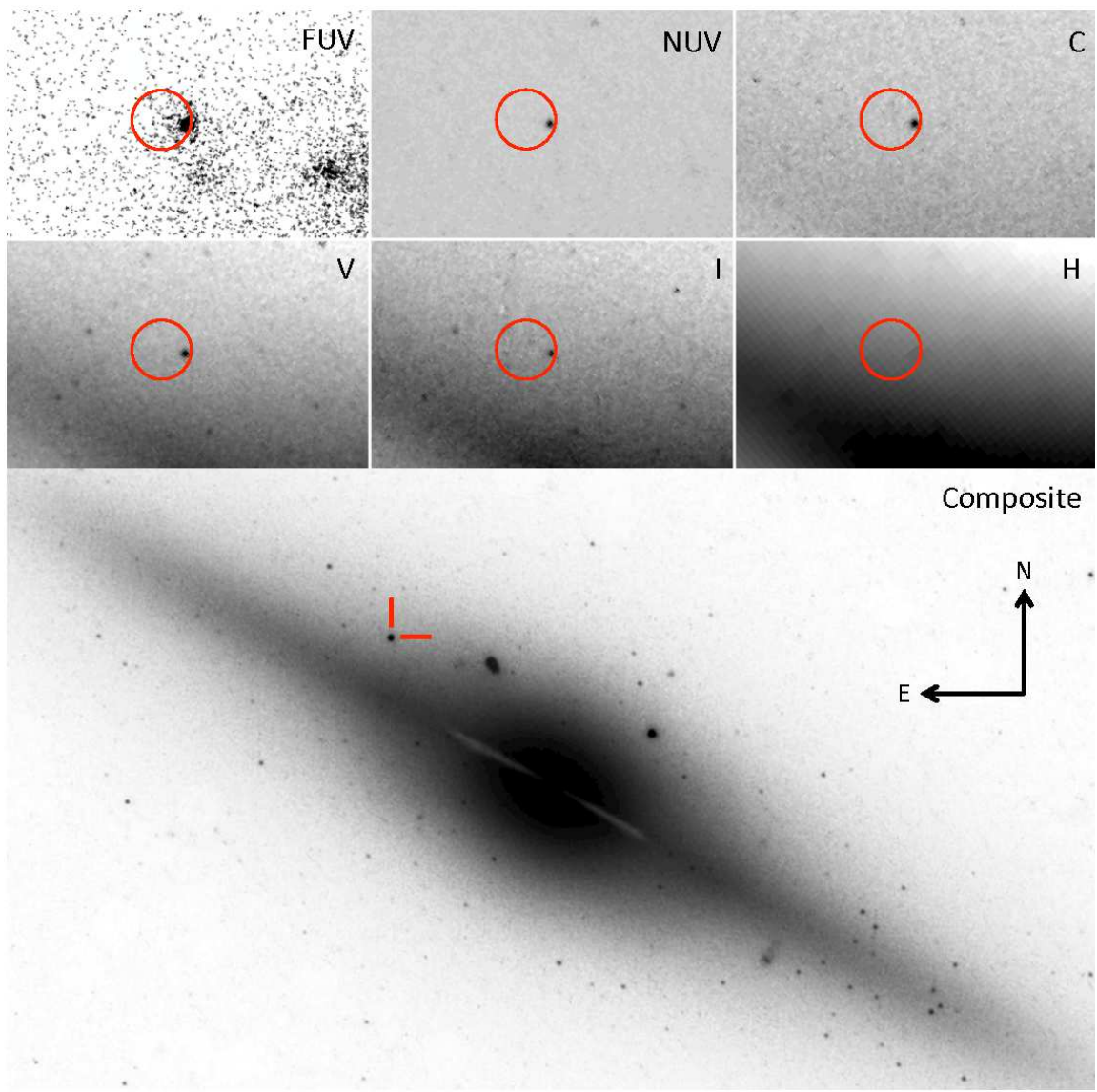}
\caption{\emph{Top panels:} \emph{HST} images from each of the 6 filters (far-UV, near-UV, Washington C-band, V-band, I-band, and H-band) zoomed in on HLX-1. The circles indicate the X-ray position of HLX-1 with the radii of 0.5\arcsec\ indicating the combined \emph{HST} plus \emph{Chandra} 95\% astrometric error. \emph{Bottom:} Composite \emph{HST} image constructed from all UV, optical and near-IR images. Prominent dust lanes around the nucleus of ESO 243-49 are evident. The HLX-1 counterpart is indicated by the tick marks. The two sources directly to the West of HLX-1 are a pair of background galaxies. \label{altimg}}
\end{figure*}

A single point source was significantly detected in all the \emph{HST} images within the 95$\%$ confidence levels of the combined \emph{Chandra} and \emph{HST} astrometry of HLX-1 (see Figure \ref{altimg}). The source is unresolved in all the images\footnote{In the FUV image it appears to be extended, but is in fact consistent with the FWHM of the larger ACS-SBC PSF.}. The 0.08$\arcsec$ FWHM of the point spread function (PSF) in the raw WFC3-UVIS images converts to a diameter upper limit of 40 pc at the distance to HLX-1 \citep[95 Mpc,][]{wie10}.



We extracted the flux of the  HLX-1 counterpart using aperture photometry with the {\tt astrolib} IDL procedure\footnote{http://idlastro.gsfc.nasa.gov} {\tt aper}, after centring with the {\tt cntrd} procedure. We used extraction radii that encircle at least 90$\%$ of the energy from the source (see Table \ref{mags}). The encircled energy curve was calculated from the Tiny Tim PSF models\footnote{http://www.stsci.edu/hst/observatory/focus/TinyTim} \citep{kri95} which consistently match the identified stars in the raw images, degraded with a Gaussian filter to reproduce the effect of image combination with drizzle. We then subtracted the background, estimated in an annulus around the source, and applied aperture corrections (see Table \ref{mags}) to the net flux to obtain the total flux from the target.


For the WFC3-IR F160W image, the galaxy emission is about 10 times
higher than the flux of the counterpart to HLX-1 and the background is
not symmetric. We therefore interpolated the extended emission of the galaxy in
a 10-pixel radius region around HLX-1 using the
procedure {\tt grid\_tps} in IDL, and subtracted it from the image in order
to obtain a flat and symmetric background around the source. We then
performed similar aperture photometry as for the other bands.


The fluxes were converted into AB magnitudes using the magnitude zero points given in the \emph{HST} WFC3 and ACS documentation and webpages\footnote{http://www.stsci.edu/hst/wfc3/phot\_zp\_lbn}\footnote{http://www.stsci.edu/hst/acs/analysis/zeropoints}. Errors on the photometry were computed with the {\tt aper} procedure as the quadratic sum of the scatter in background values, the random photon noise, and the uncertainty in mean sky brightness. The magnitudes obtained for the HLX-1 counterpart in each filter are given in Table \ref{mags}. The fluxes and errors were then converted into units of photons cm$^{-2}$ s$^{-1}$ and  {\tt XSPEC}-readable spectral files (with diagonalized response files) were generated using the  {\tt ftools} task  {\tt flx2xspec}.

\begin{table*}
\begin{center}
\caption{\emph{HST} photometry of the HLX-1 counterpart.\label{mags}}
\begin{tabular}{ccccccccc}
\tableline\tableline
Band & Filter  &  $\lambda_{pivot}$ & FWHM & \multicolumn{2}{c}{Aperture Size} &  \multicolumn{2}{c}{Encircled Energy} & Magnitude\\
 & & (\AA) & (\AA) & Src & Bkg &Src\tablenotemark{a} & Bkg\tablenotemark{b} &(AB mag)\\
\tableline
FUV & F140LP &  1527 & 294.08 & 0.5$\arcsec$ & 0.5 -- 0.6$\arcsec$ & 90.5$\%$ & 1.8$\%$& 24.11 $\pm$ 0.05 \\
NUV &F300X &  2829.8 & 753 & 0.4$\arcsec$ & 0.4 -- 0.5$\arcsec$&90.4$\%$&1.2$\%$& 23.96 $\pm$ 0.04 \\
C & F390W &  3904.6 & 953 & 0.4$\arcsec$ &0.4 -- 0.5$\arcsec$& 91.6$\%$&1.2$\%$&23.92 $\pm$ 0.06 \\
V & F555W &  5309.8 & 1595.1 & 0.4$\arcsec$ &0.4 -- 0.5$\arcsec$& 92.1$\%$&1.1$\%$&23.83 $\pm$ 0.08 \\
I &F775W &  7733.6 & 1486 & 0.3$\arcsec$ &0.2 -- 0.3$\arcsec$& 90.2$\%$&2.1$\%$& 23.91 $\pm$ 0.08 \\
H &F160W &  15405.2 & 2878.8 & 0.5$\arcsec$ & 0.5 -- 0.6$\arcsec$& 90.6$\%$&1.2$\%$& 24.4 $\pm$ 0.3 \\
\tableline
\end{tabular}
\tablenotetext{a}{Fraction of encircled energy in the aperture.}
\tablenotetext{b}{Fraction of the source energy included in the background anunulus.}
\end{center}
\end{table*}

The fitting of the broadband spectral energy distribution (SED) constructed from the \emph{HST} and \emph{Swift} data was performed using  {\tt XSPEC} v12.6.0q \citep{arn96}. The \emph{Swift} XRT data were consistent with both multi-colour disc black body and power law models, however, the power law photon index of $\Gamma$ = 4.9 is unphysically steep. We thus assumed a thermal model for the X-ray emission and fitted the SED with an irradiated disc model \citep[\emph{diskir};][]{gie08,gie09}, which includes thermal emission from the inner disc, non-thermal contribution from the Compton tail, and reprocessing in the outer disc. Components representing absorption by the neutral hydrogen column \citep[using the \emph{tbabs} model and the elemental abundances prescribed in][]{lod03} and dust extinction \citep[using the \emph{redden} model and the extinction curves in][]{car89} were included. An acceptable fit was not obtained ($\chi^2$/dof = 48.26/30), with a clear excess of UV/optical/near-IR emission evident in the residuals. 



HLX-1's position $\sim$1 kpc out of ESO 243-49's plane is naturally explained if the BH were embedded in a cluster of stars. We therefore attempted to fit the red excess in the SED using stellar population models representing emission from such a stellar cluster. We generated {\tt XSPEC} additive table models\footnote{http://www.maraston.eu/Xspec\_models} for the \citet{mar05} stellar population models (based on theoretical atmospheres with the Salpeter initial mass function) using the {\tt ftools} routine {\tt wftbmd}.
We added this component to the absorbed, reddened, irradiated disc model with no priors imposed on the age or metallicity of the stellar component. The redshift was frozen at the value of z = 0.0223 obtained from VLT spectroscopy \citep{wie10} but the remaining parameters were free to vary\footnote{The upper limit of the age was constrained to be 13.5 Gyr, consistent with the age of the Universe minus the light travel time from HLX-1 to Earth.}. 

Two distinct fits were obtained with the irradiated disc plus stellar population model. One solution had a young stellar population (13 Myr) with a low fraction of bolometric flux reprocessed in the outer disc. The second solution was with an older stellar age (13 Gyr) and high levels of disc irradiation. Ages up to $\sim$200 Myr are also allowed (giving $f_{\rm out}$ $\sim$ 1 $\times$ 10$^{-3}$), but solutions between $\sim$200 Myr -- 10 Gyr required unphysically high fractions of reprocessing in the outer disc \citep[$\gg 10\%$ of the bolometric flux, implying that the solid angle of the disc seen by the central BH is $>$ 4$\pi$ sr; see][]{gie09}. The best fit outer disc radii in both fits were $\sim$10$^3$ times the inner disc radius. Such a small disc would be stable in the context of the thermal-viscous disc instability model \citep{las11}, adding further weight to the arguments against this mechanism driving the observed outbursts. The stellar metallicity, irradiated disc fraction, and outer disc radius were poorly constrained, preventing us from drawing any strong conclusions regarding the nature of the stellar population or the size and level of irradiation of the disc. In addition, the fact that we can fit the data with two models with such disparate parameter values indicates degeneracies in the fit, particularly between the stellar age and fraction of outer disc irradiation. Nonetheless, using the age, metallicity, and luminosity of the stellar components in each fit, we are able to constrain the stellar-mass to be $\sim$(4 -- 6) $\times$ 10$^{6}$ M$_\odot$. Table \ref{specpar} lists the parameter values obtained with this fit.

\begin{table*}
\begin{center}
\caption{Best-fit spectral parameters for both the young and old stellar population solutions.\label{specpar}}
\begin{tabular}{lcccc}
\tableline\tableline
Parameter & Symbol & \multicolumn{2}{c}{Values\tablenotemark{a}} &  Units\\
& & Young & Old &\\
\tableline
Extinction & E(B $-$ V) &  0.42$^{+0.06}_{-0.4}$ & 0.2$^{+0.3}_{-0.2}$ 	&mag\\
Absorption & $N_{\rm H}$ &  0.1$^{+0.06}_{-0.04}$ & 0.04$^{+0.07}_{-0.04}$	&10$^{22}$ cm$^{-2}$\\
\tableline
\multicolumn{5}{c}{Stellar Population Component}\\
\tableline
Metallicity\tablenotemark{b} & Z$_*$ & 1 & 2 &Z$_\odot$\\
Age\tablenotemark{b} &Log(Age$_*$) & $<$ 7.1 & 10.1	& Log(yr)\\
Bolometric stellar population luminosity & L$_*$ &1.4 $\times$ 10$^{42}$ & 5.5 $\times$ 10$^{39}$& erg s$^{-1}$ \\
 &  & 3.7 $\times$ 10$^{8}$ & 1.4 $\times$ 10$^{6}$ &L$_\odot$ \\
Stellar mass & M$_*$ & 4 $\times$ 10$^6$ & 6 $\times$ 10$^6$&M$_\odot$ \\
\tableline
\multicolumn{5}{c}{Irradiated Disc Component}\\
\tableline
Disc temperature & kT$_{d}$ & 0.19$^{+0.03}_{-0.02}$& 0.21$^{+0.03}_{-0.04}$	& keV \\
Photon index\tablenotemark{c} & $\Gamma$  & 2.1   & 2.1  &\nodata\\
High-energy turn over temperature\tablenotemark{c} & kT$_{e}$ & 100  & 100  & keV\\
Ratio of Compton tail to disc luminosity & L$_c$/L$_d$ & 0.09$^{+0.1}_{-0.05}$ & 0.13$^{+0.1}_{-0.09}$ & \nodata \\
Compton inner disc fraction\tablenotemark{c} & $f_{\rm in}$  & 0  & 0 & \nodata \\
Radius of Compton illuminated disc\tablenotemark{c} & $r_{\rm irr}$ & 1.001  &1.001  & $R_{\rm in}$ \\
Fraction of flux thermalised in outer disc\tablenotemark{b} & $f_{\rm out}$ & 8 $\times$ 10$^{-7}$ & 0.098$^{+0.002}_{-0.07}$ &\nodata \\
Outer disc radius\tablenotemark{b} & Log($r_{\rm out}$) & 3.4 & 3.4$^{+0.3}_{-3}$&Log($R_{\rm in}$) \\
Bolometric disc luminosity & L$_{d}$ & 1.1 $\times$ 10$^{42}$ &  1.1 $\times$ 10$^{42}$ & erg s$^{-1}$ \\
\tableline
Fit statistics &$\chi^2$/dof &23.38/27 & 24.28/27&\nodata \\
\tableline
\end{tabular}
\tablenotetext{a}{All errors are quoted at the 90$\%$ confidence level.}
\tablenotetext{b}{These parameters without errors could not be constrained. For example, values as high as 1 $\times$ 10$^{-3}$ and 6 are allowed for the $f_{\rm out}$ and $r_{\rm out}$ parameters respectively in the young solution.}
\tablenotetext{c}{Contribution from a Compton component is minimal, so we froze kT$_{e}$ = 100 keV, $\Gamma$ = 2.1 \citep[consistent with deeper observations of HLX-1 and stellar-mass BH binaries in the same luminosity state][]{sev11,don07}, $f_{\rm in}$ = 0 and $r_{\rm irr}$ = 1.001, setting the Compton illumination component to zero.
}
\end{center}
\end{table*}

\begin{figure*}
\begin{center}
\includegraphics[width=\textwidth]{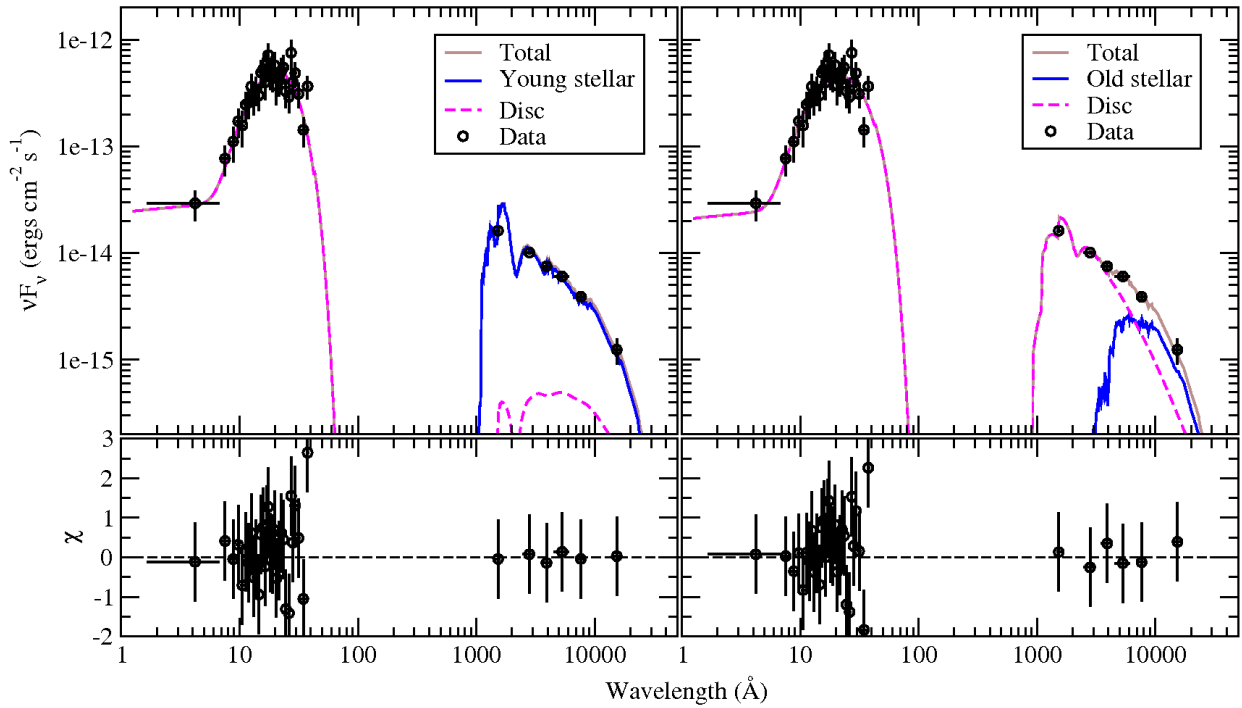}
 \caption{Best-fit broadband SED model of HLX-1 constructed using \emph{Swift} XRT and \emph{HST} data. \emph{Left:} fitted with a model representing low level disc irradiation plus a young stellar population model. \emph{Right:} fitted with a model representing a highly irradiated disc plus an old stellar population. The bottom panels show the fit residuals.}\label{sed}
\end{center}
\end{figure*}

To test for model dependency in the fit, we also generated additive table models using stellar population models based on empirical stellar libraries by \citet{mar11}. In particular, we used the models based on the \citet{pic98} Solar metallicity library, because of their extended wavelength range. Replacing the theoretical atmosphere models with these empirical stellar library models also obtained fits with both young ($\chi^2$/dof =  23.81/28) and old ($\chi^2$/dof =  24.52/28) stellar populations.


\section{Discussion \& Conclusions}

The detection of a stellar population around HLX-1 provides insights into the origin of and environment around an IMBH. Our SED fitting could not constrain the age of the stellar cluster around HLX-1 other than to show that only young or old populations are allowed. However, the reprocessing fraction in the outer disc required for old solution is $\sim$10$\%$, which borders on being non-physical \citep[reprocessing fractions $>$10$\%$ require the disc to subtend an unfeasibly large solid angle;][]{gie09}. In addition, the size of the accretion disc is constrained to be extremely small with an outer radius $\sim$1000 times the inner disc radius. 

Galactic low mass X-ray binaries in disc-dominated states typically have fractions of reprocessing in the outer disc of $f_{\rm out}$ $\sim$ 10$^{-3}$ \citep{gie09}, far lower than the value obtained in the old stellar population solution. The outer disc radius also has a strong effect on the reprocessing fraction. Freezing the outer disc radius at a more likely value of 10$^5$$R_{\rm in}$ (C. Done 2011, private communication) we could not find an acceptable fit solution with an old population ($\chi^2$/dof = 225.42/29), indicating that an extremely high reprocessed fraction and an extremely small disc are necessary outcomes with an old star cluster\footnote{In contrast, the poor constraints on the outer disc radius in the young stellar population solution allow for a much larger value up to 10$^6$$R_{\rm in}$ without an increase in the $\chi^2$.}. We therefore favour the young stellar population solution, although without additional data we cannot conclusively rule out an extremely old star cluster. The upper limit of 40 pc derived for the cluster diameter is consistent with either a globular cluster or young massive star cluster, which typically have half-mass radii of $\sim$10 pc \citep{har96} and $<$ 50 pc \citep{por10}, respectively.

The presence of a young stellar population is difficult to reconcile with HLX-1 residing in a classical globular cluster (such as those observed in our own Galaxy), which are typically dominated by old stars \citep[e.g.][]{for03}. However, globular clusters with young stellar populations have been observed around disrupted galaxies such as the Antennae \citep{bas06} and the Magellanic Clouds \citep{els85}. The mass of the cluster around HLX-1 (calculated using the derived age and metallicity, the observed luminosity and the model mass-to-light ratio\footnote{See http://www.maraston.eu}) is $\sim$4 $\times$ 10$^{6}$ M$_\odot$, which is at the upper end of the standard classical globular cluster mass range \citep{mar04}. 

A young star cluster surrounding HLX-1 could also be explained in the accreted dwarf galaxy scenario \citep{kni10}. Tidally stripping a dwarf galaxy during a merger event could remove a large fraction of the mass from the dwarf galaxy, with star formation triggered as a result of the tidal interactions. This could result in the observed IMBH embedded in the remnant of the nuclear bulge and surrounded by a young, high metallicity, stellar population. It has been proposed that such accreted dwarf galaxies may explain the origin of some globular clusters, with the remnant cluster appearing more like a classical globular cluster as its stellar population ages \citep{for10}. 

A link has been drawn between the presence of prominent dust lanes in early-type galaxies with frequent gas-rich minor mergers, with the host galaxy nuclear BH becoming active within $<$ 200 Myr following the merger event \citep{sha11}. The \emph{HST} images of ESO 243-49 (see Figure \ref{altimg}) reveal pronounced dust lanes and yet no evidence of nuclear activity was detected in \emph{Chandra} X-ray images \citep{sev11}, implying that the merger events that contributed to the formation of the dust lanes took place in the recent past. This is thus consistent with the presence of a young stellar population surrounding HLX-1, where localised star formation would have been triggered as a result of tidal interactions with ESO 243-49 following a recent merger event. 


We note that \citet{sor11} argue against a young stellar population using lower signal-to-noise UBVR data from the Very Large Telescope. However, they do not fit the X-ray and optical data simultaneously and use alternative stellar population models. Additional \emph{HST} observations obtained at different X-ray luminosities are thus required in order to test the competing theories and therefore constrain the disc irradiation and stellar population contributions.

\acknowledgments

We thank the referee for useful comments and J.-P. Lasota, T. Alexander, C. Done, S. Shabala, F. Gris\'{e}, K. Arnaud and R. Starling for helpful discussions. SAF is the recipient of an ARC Postdoctoral Fellowship, funded by grant DP110102889. SAF, TJM, AJG and KW acknowledge funding from the UK STFC. MS acknowledges support from NASA grants DD0-11050X andÊGO 12256. JP and CM acknowledge the Marie-Curie Excellence Team grant ÔUnimassÕ, ref. MEXT-CT-2006-042754 of the TMR programme financed by the European Community. NAW, DB and OG thank the CNES. Based on observations made with the NASA/ESA \emph{HST}, obtained at the STScI, which is operated by the Association of Universities for Research in Astronomy, Inc., under NASA contract NAS 5-26555. These observations are associated with program \#12256. Also based on observations with \emph{Swift}. 



{\it Facilities:} \facility{HST (ACS)}, \facility{HST (WFC3)}, \facility{Swift (XRT)}, \facility{CTIO}.

\clearpage

\end{document}